\UseRawInputEncoding
\documentclass[conference]{IEEEtran}

\ifCLASSINFOpdf
\else
\fi
\usepackage{amsmath}
\usepackage{amssymb}
\usepackage{algorithmic}
\usepackage{graphicx}
\usepackage{array}
\usepackage{fixltx2e}
\usepackage{stfloats}
\usepackage{cite}
\usepackage{url}
\DeclareMathAlphabet{\mathpzc}{OT1}{pzc}{m}{it}

\begin{document}
\title{Comparison of different coding\\ schemes for 1-bit ADC}
\author{\IEEEauthorblockN{Fedor Ivanov}
\IEEEauthorblockA{National Research University\\Higher School of Economics\\
Moscow, Russia\\
Email: fivanov@hse.ru}
\and
\IEEEauthorblockN{Dmitry Osipov}
\IEEEauthorblockA{National Research University\\Higher School of Economics,\\
Institute for Information\\
Transmission Problems\\
Moscow, Russia\\
Email: d\_osipov@iitp.ru}}

\maketitle
\begin{abstract}
This paper devotes to comparison of different coding schemes (various constructions of Polar and LDPC codes, Product codes and BCH codes) for the case when information is transmitted over AWGN channel with quantization with lowest possible complexity and resolution:  1-bit. We examine performance (in terms of Frame-error-rate --- FER) for schemes mentioned above and give some reasoning for results we obtained. Also we give some recommendations for choosing coding schemes for a given code rate and code length.
\end{abstract}
\IEEEpeerreviewmaketitle

\section{Introduction}
 In modern communication system each receiver antenna is generally
equipped with a dedicated radio-frequency (RF) chain that includes complex, power-hungry  analog-to-digital converters (ADCs). Even though it is known that the transmit power can be made inversely proportional to the number of antennas, the power dissipated by each ADC scales linearly with the sampling rate (baseband bandwidth) and exponentially with the number of quantization bits. High resolution ADCs also increase  receiver complexity. Therefore as the number of antennas used by communication system grows the use of low resolution ADCs seems more and more appealing. The most extreme form of a low resolution ADC is a 1-bit ADC which has low implementation complexity since it includes a single comparator. Another important benefit of a 1-bit ADC is the fact that the analog stages of the RF chain (Automatic Gain Control (AGC), mixer, and analog filters)  can be discarded or converted to digital parts. Thus 1-bit ADC is very promising for power and complexity-constrained communication systems such as sensor networks and IoT \cite{Varasteh} and even more promising for MIMO systems especially due to the constant growing popularity of massive MIMO.

Although considerable amount of efforts has been spent recently to develop advanced signal processing techniques( e.g. channel estimation \cite{Risi,Choi,Li,Muta} and equalization \cite{Mohamed}) for receivers with 1-bit ADCs little is known about error-correction coding for communication systems employing 1-bit ADCs. Most research papers on the topic published recently deal with MU MIMO and massive MIMO scenarios \cite{Kim,Lee,Cho,Amrallah} and/or additional processing techniques such as oversampling \cite{Alencar}, successive cancellation \cite{Cho}, channel-based precoding \cite{Kim} or deep learning-based autoencoders \cite{Balevi,Andrews}. This paper concentrates on a more general and simple problem setting that is not restricted to certain class of communication systems or advanced signal processing techniques.
\section{Channel Model}

Let us first describe channel model we use. 

We assume that information vector ${\bf c}\in\{0,1\}^n,\:n\in\mathbb{N}$ modulated by BPSK modulator (mapped to ${\bf x}=\{-1,+1\}^n$ by the rule: $x_i=(-1)^{c_i}$) is transmitted over channel with additive white Gaussian noise (AWGN). Thus receiver obtains:
$$
{\bf y}={\bf x}+{\bf \eta},
$$
where $\eta=(\eta_1,\:\eta_2,\ldots,\eta_n)$, and $\eta_i\sim \mathcal{N}(0,\sigma^2)$, $i=1..n$ are pairwise independent and identically distributed (by normal distribution) random variables. Under $\sigma^2$ we denote noise variance. Thus each $\bf y$ is vector of independent random variables distributed either by $\mathcal{N}(1,\sigma^2)$ (if $x_i=1$) or by $\mathcal{N}(-1,\sigma^2)$ (if $x_i=-1$).

We assume that received signal $\bf y$ is quantized with lowest possible $1$-bit resolution (1-bit ADC). In this paper we deal with only symmetric case and thus quantization rule $\mathcal{Q}:{\bf y}\mapsto{\bf q}$ can be expressed as follows:
$$
q_i=\mathrm{sign}(y_i),
$$
It is clear that composite channel AWGN + 1-bit symmetric ADC is equivalent to binary symmetric channel (BSC) with transition probability \cite{Singh}:
$$
p=Q(\sqrt{SNR}), 
$$
where $Q(x)=\frac{1}{\sqrt{2\pi}}\int\limits_x^{\infty}e^{-\frac{t^2}{2}}dt$ is the complementary Gaussian distribution function and SNR is a signal-noise ratio that can be calculated from noise variance $\sigma^2$ as 
$$
SNR=-10\log_{10}\sigma^2.
$$
Thus capacity of the channel we use is:
$$
C=1-h(p), 
$$
where $h(p)=-p\log_2p-(1-p)\log_2(1-p)$ is a binary entropy function.

\begin{figure}[h]
\centering
\includegraphics[width=3.2in]{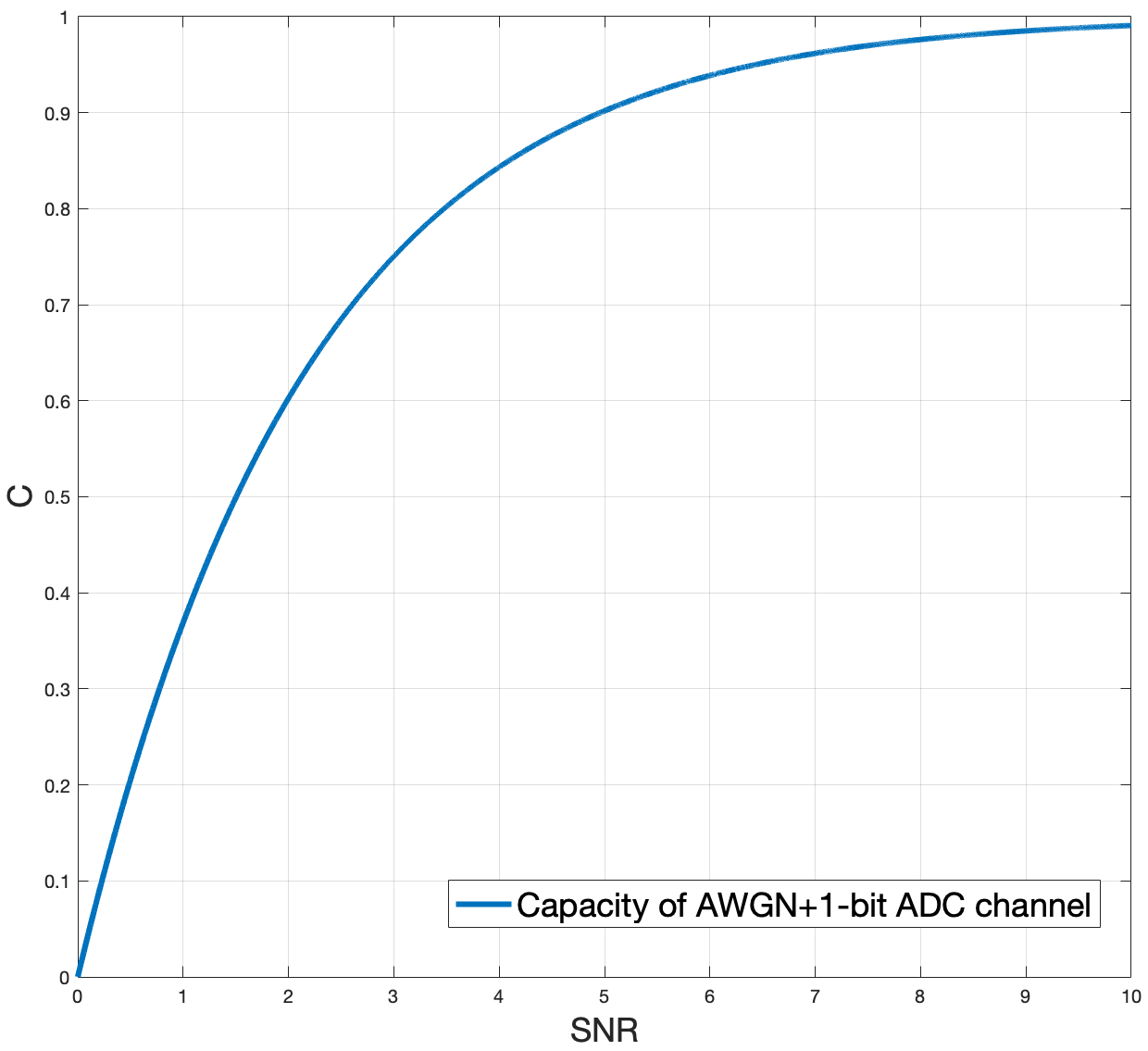}
\caption{Dependence between SNR in Gaussian channel and capacity of the AWGN+1-bit ADC channel.}
\label{cap}
\end{figure}

Fig.~\ref{cap} represents dependence between SNR in Gaussian channel and capacity of the AWGN+1-bit ADC channel.

\section{Codes for 1-bit ADC}
In this section we give a brief description of different coding schemes. We further compare the performance of these codes (in terms of frame-error-rate FER versus SNR) in AWGN+1-bit ADC channel. As a codes-candidates we consider Polar codes, Low-Density Parity-Check (LDPC) codes, Bose-Chaudhuri-Hocquenghem Codes (BCH codes) and Product codes also known as Turbo Product Codes (TPC).

For different decoders of these codes we consider 2 alternative outputs of channel. If decoder operates with ''hard'' inputs (BCH codes, Product BCH codes) then we demodulate $q_i$ and obtain bit values $r_i$ as follows $r_i=\frac{1-q_i}2$. If decoder (LDPC codes, Polar codes) works with soft information (log-likelihood ratios) obtained from the channel then input will be: $r_i=q_i\ln\frac{1-p}{p}$, where $p$ is the transition probability of BSC channel.

\subsection{Polar Codes}

Hereinafter we use standard notation: for any vector ${\bf v}=(v_0,...,v_{N-1})$ under ${\bf v}_i^j$ we understand $(v_i,v_{i+1},...v_j)$.

Polar codes were proposed by Arikan in \cite{Arikan}.

Consider a binary $(N,K)$ polar code specified by set $\mathcal{I}$ of information indexes, $|\mathcal{I}|=K$, $N=2^n,\:n\in\mathbb{N}$ and the corresponding encoding procedure:
\begin{equation}
\mathbf{x}_0^{N-1} = \mathbf{d}_0^{N-1}\mathbf{G}_N,
\label{eEnc}   
\end{equation}
where $\mathbf{d} \in \{0,1\}^N$ and $\mathbf{G}_N$ is the generator matrix of order $N$, defined as $\mathbf{G}_N = \mathbf{F}^{\otimes n} $ with the Arikan's standard polarizing kernel $\mathbf{F}\triangleq \left[ \begin{smallmatrix}
    1 & 1\\  
    0 & 1\\
        \end{smallmatrix}
\right]$ and $\otimes n$ is $n$-th Kronecker product of $\bf F$.

Considering the information $\mathcal{I}$ and frozen set $\mathcal{F}=\mathcal{I}^c$, we may write \eqref{eEnc} as
\begin{equation*}
\mathbf{x}_0^{N-1} = \mathbf{d}_{\mathcal{I}}\mathbf{G}_N(\mathcal{I})\oplus \mathbf{d}_{\mathcal{F}}\mathbf{G}_N(\mathcal{F}).  
\end{equation*} 

We consider the frozen bits as zeroes, $\mathbf{d}_{\mathcal{F}} = \mathbf{0}$, and the information bits as the information to be encoded, $\mathbf{d}_{\mathcal{I}} = \mathbf{u}$.

It is clear that for fixed $N$ and $K$ polar code is completely defined by the set of it's information bits $\mathcal{I}$. Information set plays crucial role in performance of polar code. There are different approaches to construct $\mathcal{I}$. In this paper we consider four of them:
\begin{itemize}
    \item 5G NR Polar (5G Polar) code is defined for any $1\leq K<N\leq 1024.$ This code has nested construction of information set. More details can be found in \cite{Bioglio}.
    \item Polarization weight (PW) Polar codes (PW Polar) \cite{Huawei} gives the reliability ordering as a function of their indices, i. e. this method is channel and SNR independent.
    \item Information set of Reed-Muller like Polar code (RM Polar) consists of $K$ indexes that corresponds to $K$ columns/rows of matrix $\mathbf{G}_N$ with $K$ highest Hamming weights.
    \item Gaussian approximation (GA) technique \cite{Trif} is another approach to construct set of information indexes $\mathcal{I}$. GA is used to evaluate the reliability of channels and then $K$ channels with highest reliabilities are chosen as information set (GA Polar). GA uses a two-segment approximation function to cut-off the computational complexity of Density Evolution (DE) \cite{Mori} when applied to binary input additive white Gaussian noise (AWGN) channels, but yielding almost the same precision. 
\end{itemize}

Polar codes are usually decoded either by Successive Cancellation (SC) decoding \cite{Arikan} or by Succesive Cancellation List Decoding (SCL) decoding \cite{Tal} that maintains $L$ concurrent decoding candidates.

\subsection{LDPC Codes}

Low-density parity-check codes were proposed by R.G. Gallager (G-LDPC codes) in \cite{Gallager}. The performance of G-LDPC codes for the binary symmetric channel (BSC) were studied in \cite{Chilappagari}.

It is convenient to specify LDPC codes using their Tanner graph representation \cite{Tanner81}. The Tanner graph is a bipartite graph, where the nodes on the left side are associated with the codeword bits (variable nodes) and the nodes on the right are associated with the parity-check equations (check nodes). Any LDPC code can be described in terms of bipartite graphs that are characterized by two probability vectors
\[
\tilde{{\mathbf{\lambda }}} = \left( {{\tilde{\lambda} _2}, \ldots ,{\tilde{\lambda} _c}} \right),
\]
\[
\tilde{{\bf{\rho }}} = \left( {{\tilde{\rho}_1}, \ldots ,{\tilde{\rho}_d}} \right),
\]
where $\tilde{\lambda}_{l}$ is the fraction of variable nodes with the degree $l$, and $\tilde{\rho}_{l}$ is the fraction of check nodes with the degree $l$. For convenience we also define the polynomials
\[
\tilde{\lambda} \left( x \right) = \sum\limits_{l = 2}^c {{\tilde{\lambda} _l}{x^{l - 1}}},
\]
\[
\tilde{\rho} \left( x \right) = \sum\limits_{l = 2}^d {{\tilde{\rho} _l}{x^{l - 1}}}.
\]

There are different approaches to construct LDPC codes (i. e. to specify polynomials $\tilde{\lambda} \left( x \right)$ and $\tilde{\rho }\left( x \right)$).  One group of methods is based on progressive edge growth algorithm (PEG). This algorithm aims to construct Tanner graphs with large girth by progressively establishing edges or connections between symbol and check nodes in an edge-by-edge manner. 

Another approach starts from some small matrix ${\bf H}_{core}$ of size $m_c\times n_c$ which is known as core matrix. This matrix can be found by exhaustive search or by applying Density Evolution \cite{Richardson} to find protograph with lowest possible decoding threshold. Then each non-zero element of core matrix is substituted by $m\times m$ matrix ${\bf P}_{ij}$, $i=1..m_c$, $j=1..n_c$ (usually permutation matrices or circulants, zeros in ${\bf H}_{core}$ is substituted by $m\times m$ zero matrix) to obtain LDPC code for a desired length $n=mn_c$ with rate $R=1-\frac {m_c}{n_c}$. Usually non-zero matrices are chosen to optimize some parameter of resulting LDPC code. For instance ACE algorithm \cite{ACE} is designed to choose such circulant matrices that optimizes (decreases the number of short cycles) cycle spectrum (ACE spectrum) of resulted LDPC code.

\subsection{BCH Codes}

Binary BCH codes  is a class of cyclic algebraic codes defined by it's generator polynomial $g(x)$. A BCH code over a finite field $\mathbb{F}_2$ of length $n$ and designed distance $\delta$ is the cyclic code with generator polynomial that has $\delta-1$ consecutive roots $\alpha^b,\:\alpha^{b+1},\ldots,\alpha^{b+\delta-2}\in\mathbb{F}_2^m$, where $\alpha$ is some element of $\mathbb{F}_2^m$. In this case $n$ is a multiplicative order of $\alpha$. Important special cases are $b = 1$ (called narrow-sense BCH codes), or $n=2^m-1$ (called primitive BCH codes). In this paper BCH code is assumed to be narrow-sense and primitive.

BCH codes are usually decoded by well-known Berlekamp-Massey decoder (BM)\cite{Massey}. It is able to correct up to $\frac{d-1}2$ errors, where $d\geq\delta$ is a real minimal distance of BCH code. 

\subsection{Product Codes}

The concept of product codes (or turbo product codes - TPC) is a simple and efficient method to construct powerful and long codes with a large minimum Hamming distance, $\delta$, using conventional linear block codes \cite{Elias}. 

Let us now consider the product code construction in more detail. Consider two systematic linear block codes $C_A$ and $C_B$ having parameters $(n_A,k_A,d_A)$ and $(n_B,k_B,d_B)$ respectively. The product code $P=C_A\times C_B$ is obtained by placing $k_Ak_B$ information bits in a matrix of $k_A$ rows and $k_B$ columns and encoding the $k_A$ rows and $k_B$ columns using codes $C_A$ and $C_B$ respectively. Furthermore, the parameters of the resulting product code $P$ are given by $n_P=n_An_B$, $k_P=k_Ak_B$, the code rate $R_P$ is given by $R_P=R_AR_B$ and minimal distance $\delta_P=d_Ad_B$. Thus, it is possible to construct powerful product codes based on linear block codes such as BCH codes and other ones.

Turbo product codes usually decoded by iterative soft-input soft-output (SISO) decoders like As indicated by Elias \cite{Elias}, a TPC can be decoded by sequentially decoding it component codes in order to reduce decoding complexity. 

Unfortunately in the case of 1-bit ADC it is impossible to obtain reliable soft outputs (LLRs) from the channel and thus the only way to decode TPC is to apply hard-input decoders. Moreover, it is known that such powerful codes like LDPC have excellent performance only in the case of lengths from several thousand and thus can not be used to construct TPC with moderate length. It is also known that product of two polar codes result in new polar code with not optimal frozen set (in terms of performance over channel with binary input) \cite{Condo} and thus it would be better to consider long conventional polar code to obtain better performance. As a summary, in this paper we consider a BCH product codes with algebraic decoding of it's components.

\section{Simulation Results}

In this section we present a competitive comparison of code constructions described above. Let us first give a description of the simulation setup we provide. We also fix such code parameters as rate and length. After providing comparison we discuss these results both to explain them and choose most promising code candidates for AWGN + 1-bit ADC transmission scenario.

\subsection{Simulation Setup $N\approx 1024$}

We consider the following code parameters:
\begin{itemize}
    \item Code length: $N\approx1024$ (we can not fix one code length for all constructions since this length is unachievable for some codes (for instance for product BCH codes). If some code can not achieve exact code length then we choose the closest possible code length.
    \item Code rates: we consider a set of code rates: $R\in\{0.5, 0.625, 0.75, 0.8125, 0.875, 0.9375\}$. If some code can not achieve exact code rate (for instance BCH codes does not exist for any rate) then we choose the closest possible code rate.
\end{itemize}

We examine the following coding schemes:
\begin{itemize}
    \item LDPC codes based on DE+ACE (density evolution serves to obtain small protograph with optimal threshold and then ACE is used to extend this protograph maximizing girth)  and based on PEG (obtain optimal distribution for fixed degree of variable nodes $l=3$ for rates $0.5, 0.625, 0.75, 0.8125$ and $l=4$ for rates $0.875, 0.9375$). LDPC codes is decoded by Sum-Product decoding with $50$ iterations.
    \item Polar codes constructed by different techniques: 5G NR, PW, GA, RM (see section II). These codes are decoded by SCL decoder with list size $32$. CRC length is $16$ for all cases.
    \item BCH codes with the following parameters: $(1023,512)$,$(1023,638)$, $(1023,768)$, $(1023,828)$, $(1023,893)$ and $(1023,953)$. These codes are decoded by Berlekamp-Massey decoder.
    \item Product codes constructed from extended BCH codes and primitive BCH codes. As a result we obtain the following parameters: 
    \begin{itemize}
        \item $(32,26)\times(32,21)=(1024,546)\:(R\approx 0.5)$ 
        \item $(16,11)\times(64,57)=(1024,627)\: (R\approx 0.625)$
        \item $(64,51)\times(16,15)=(1024,765)\: (R\approx 0.75)$
        \item $(31,26)\times(32,31)=(992,806)\:(R\approx 0.8125)$
        \item $(8,7)\times(128,127)=(1024,889)\:(R\approx 0.875)$
        \item $(32,31)\times(32,31)=(1024,961)\: (R\approx 0.9375)$.
    \end{itemize}
    
    These codes decoded iteratively with $10$ iterations.
\end{itemize}

As it was mentioned above as a communication channel we consider BI-AWGN and symmetric 1-bit ADC in receiver side. Modulation scheme is BPSK.

\subsection{Comparison of Different Coding Schemes}

In this section we present the simulation results for different coding schemes described above. At each figure we compare the performance (in terms of Frame Error Rate - FER) versus SNR for different coding schemes for a fixed rate. 

\begin{figure}[h]
\centering
\includegraphics[width=\columnwidth,height=150pt]{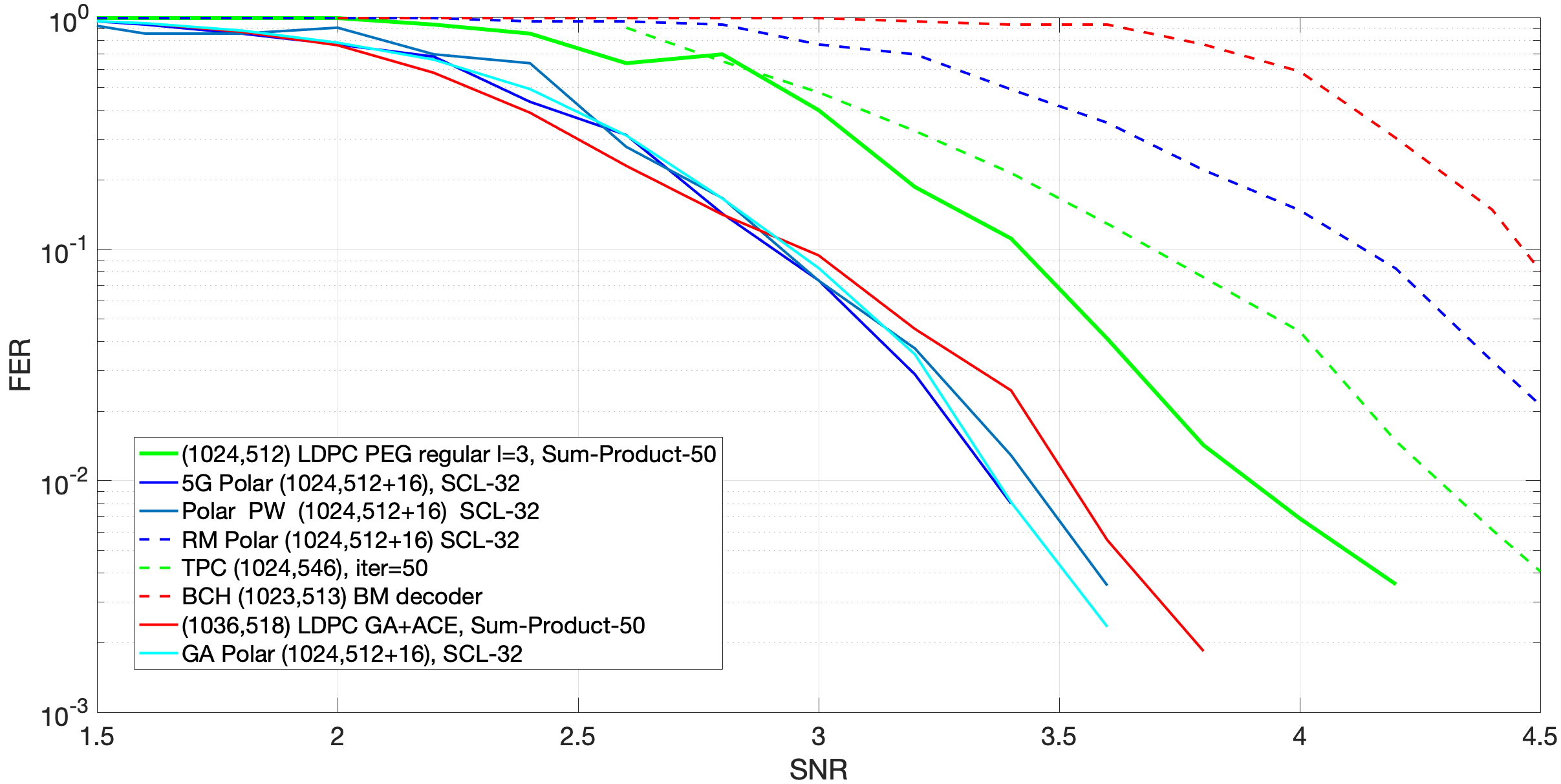}
\caption{FER versus SNR of AWGN+1-bit ADC channel for different code constructions with $N\approx 1024$, $R\approx 0.5$}
\label{comp05}
\end{figure}

Simulation results for $N\approx 1024$ and $R\approx 0.5$ are presented in Fig.~\ref{comp05}. From this figure we can see that four coding schemes (Polar 5G NR, Polar PW, Polar GA --- obtained from Density Evolution for $SNR=2$ and LDPC codes designed from small protograph obtained by density evolution give approximately the same simulation results. LDPC code that was optimized only in terms of girth in protograph (PEG LDPC) has worse performance due to the fact that only girth metric is adequate for pure AWGN channel without any quantization. Product codes (TPC) has even worse performance due the  fact that short constituent codes of TPC are BCH codes that are not able to correct enough fraction of errors for this transmission conditions. RM polar code that optimizes minimal distance of polar code also has an unsatisfactory performance since in this SNR ratio spectrum of the code is more important than minimal distance. Finally, primitive narrow sense BCH code has the worst performance --- this code is decoded by Berlekamp-Massey decoder and can not provide good performance at least for a given SNR range and code rate. 

\begin{figure}[h]
\centering
\includegraphics[width=\columnwidth,height=150pt]{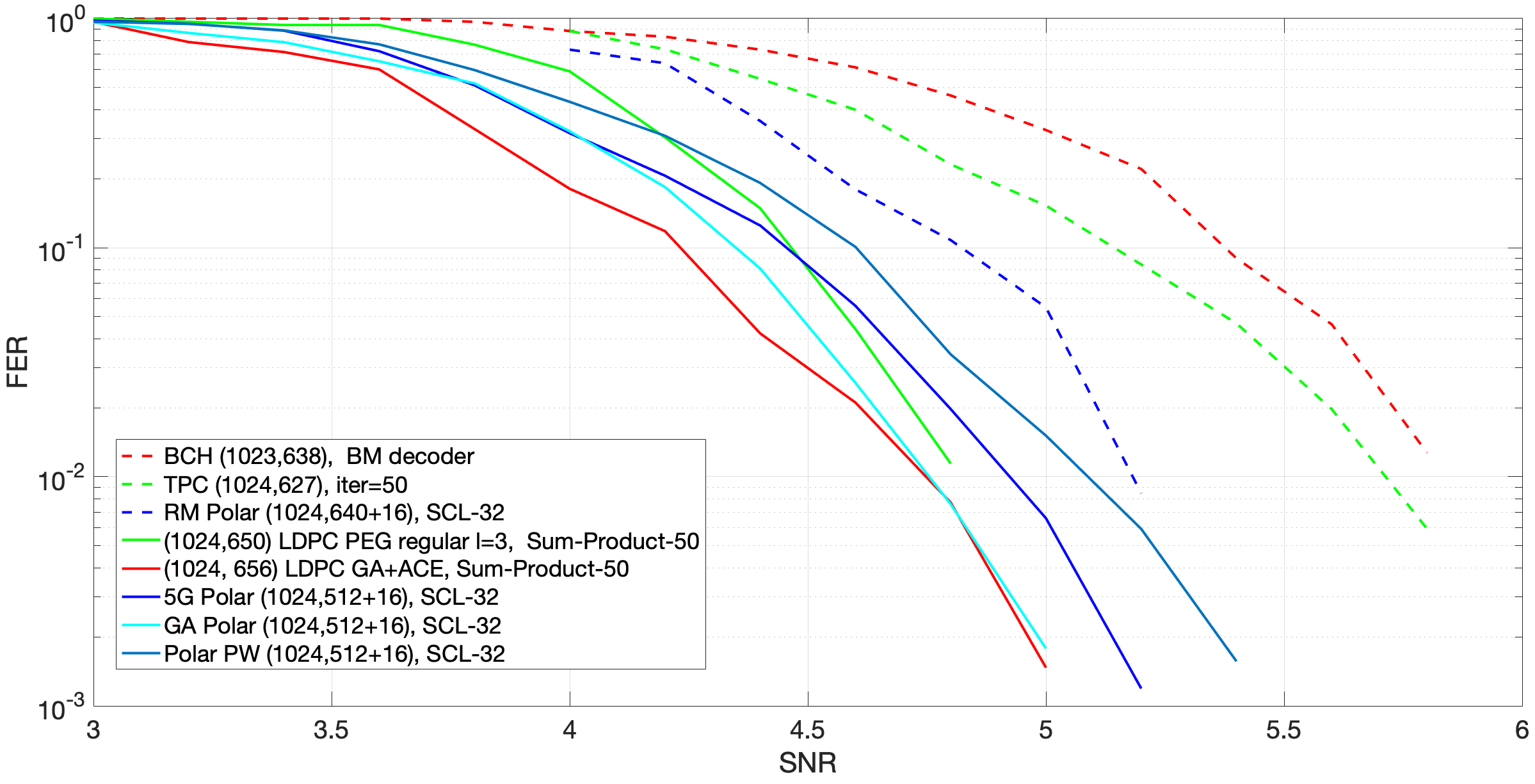}
\caption{FER versus SNR of AWGN+1-bit ADC channel for different code constructions with $N\approx 1024$, $R\approx 0.625$}
\label{comp0625}
\end{figure}

Simulation results for $N\approx 1024$ and $R\approx 0.625$ are presented in Fig.~\ref{comp0625}. From this figure we can also notice that the same four coding schemes (Polar 5G NR, Polar GA --- obtained from Density Evolution for $SNR=3.5$, DE LDPC and PEG LDPC) give approximately the same simulation results. But the best schemes are LDPC (DE  + ACE extention) and Polar DE codes.  PW Polar code has a bit worse performance due to the fact that these codes constructed in channel independent method.  RM polar code now  has comparable performance (but a bit worse) than schemes mentioned above. The reason of improving performance is the fact that the higher SNR the more important the minimal distance of the code (key metric in RM codes construction). Product codes (TPC) has even worse performance due the  fact that short constituent codes are not able to correct enough fraction of errors for this transmission conditions. Finally, primitive narrow sense BCH code has the worst performance. 

\begin{figure}[h]
\centering
\includegraphics[width=\columnwidth,height=150pt]{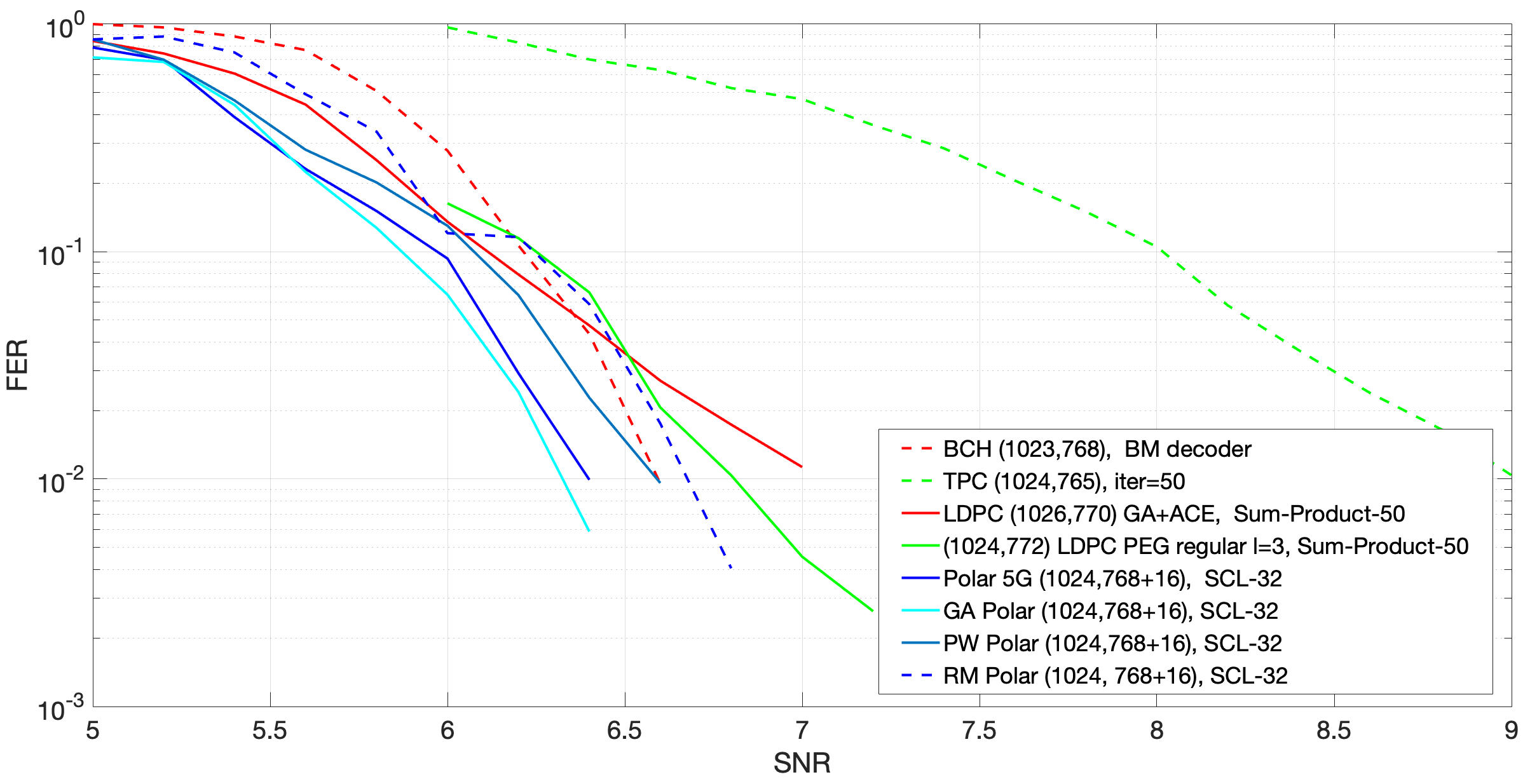}
\caption{FER versus SNR of AWGN+1-bit ADC channel for different code constructions with $N\approx 1024$, $R\approx 0.75$}
\label{comp075}
\end{figure}

Simulation results for $N\approx 1024$ and $R\approx 0.75$ are presented in Fig.~\ref{comp075}. From this figure we can see that nearly all coding schemes (but TPC codes) have approximately the same performance. The slightly better performance have 5G and DE Polar codes but the difference between DE Polar and LDPC DE + ACE code is negligible. Only TPC codes have  significantly worse performance than all other schemes because for this coding rate (and  all higher rates) it is  rather difficult to optimize inner and outer codes because these  codes are algebraic (that limits space of their possible parameters)  and rather short. Speaking about the closeness other results to each other we  can  conclude that for this coding rate BCH codes significantly improve their performance and other constructions (methods of their designing) can still result in good performance.

\begin{figure}[h]
\centering
\includegraphics[width=\columnwidth,height=150pt]{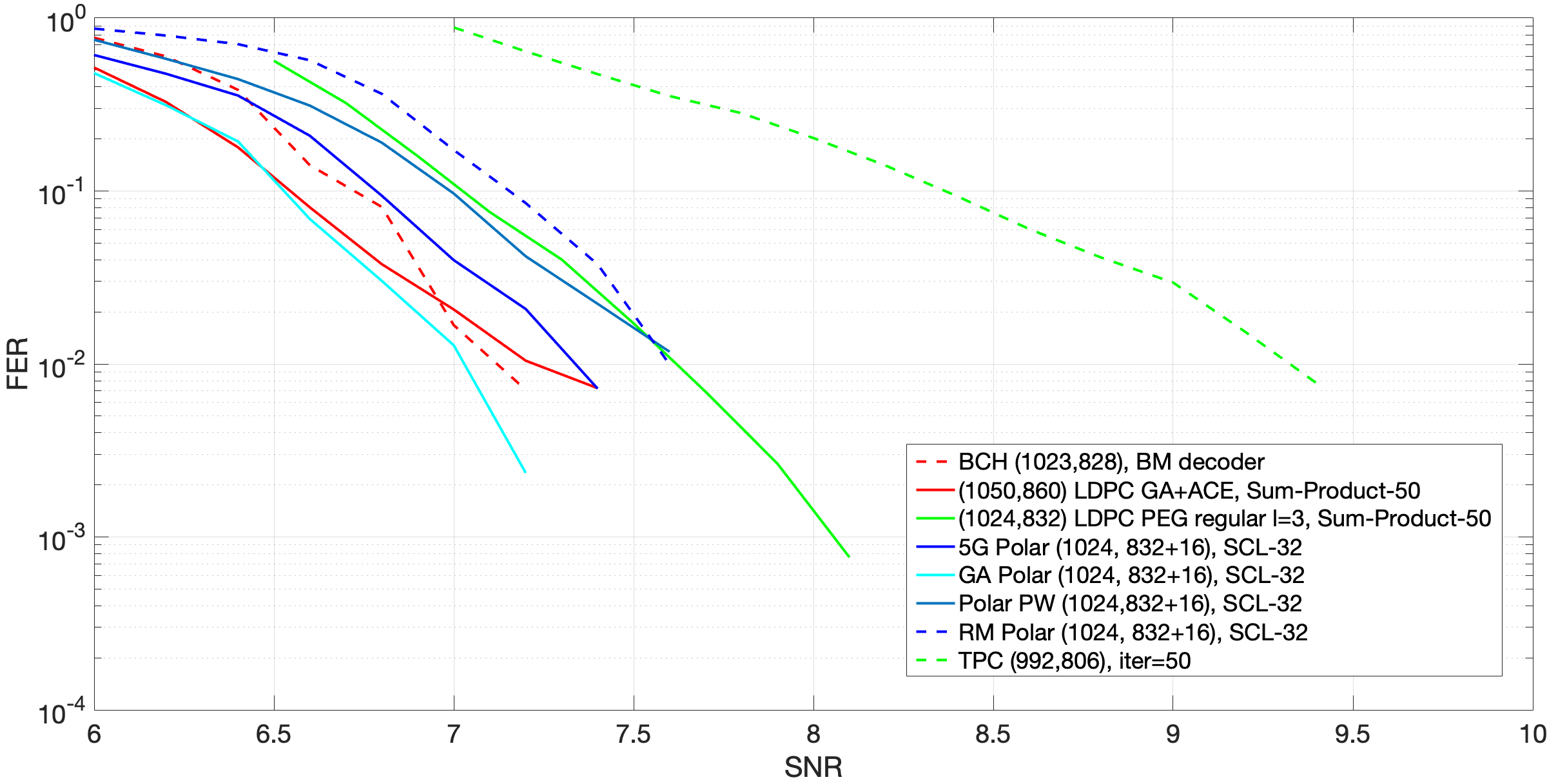}
\caption{FER versus SNR of AWGN+1-bit ADC channel for different code constructions with $N\approx 1024$, $R\approx 0.8125$}
\label{comp08125}
\end{figure}

Simulation results for $N\approx 1024$ and $R\approx 0.8125$ are presented in Fig.~\ref{comp08125}. From this figure we can see that as for $R\approx 0.8125$ nearly all coding schemes (but TPC codes) have approximately the same performance. But instead of the results for $R\approx 0.75$ now we have a new leaders: slightly better performance have 5G Polar codes, LDPC (DE + ACE extension) and BCH codes. Only TPC codes have  significantly worse performance as in previous case. 

\begin{figure}[h]
\centering
\includegraphics[width=\columnwidth,height=150pt]{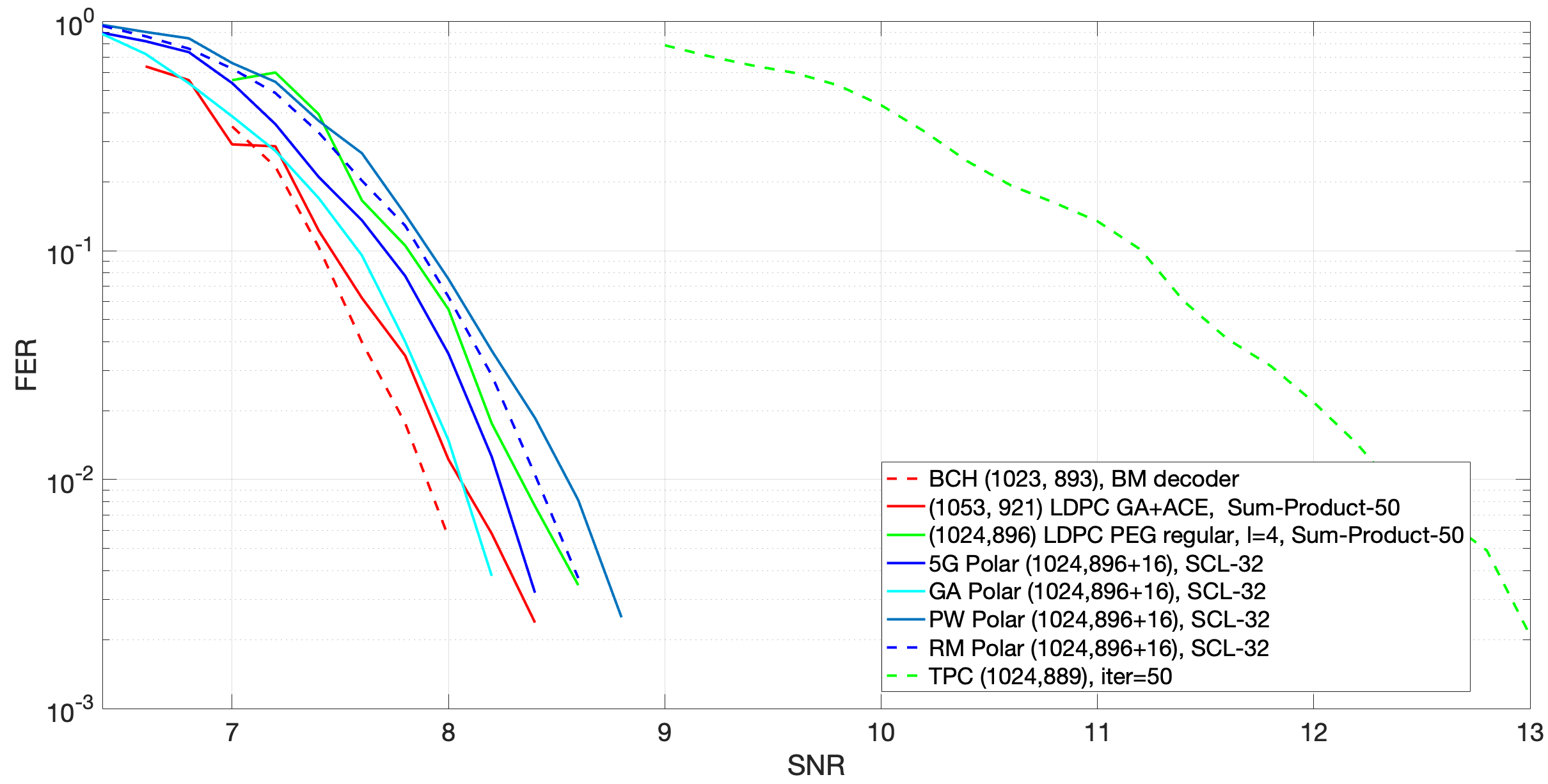}
\caption{FER versus SNR of AWGN+1-bit ADC channel for different code constructions with $N\approx 1024$, $R\approx 0.875$}
\label{comp0875}
\end{figure}

Simulation results for $N\approx 1024$ and $R\approx 0.875$ are presented in Fig.~\ref{comp0875}. From this figure we can see that as for $R\approx 0.75$ and $R\approx 0.8125$ nearly all coding schemes (but TPC codes) have approximately the same performance. But instead of the results for $R\approx 0.8125$ the best performance has BCH code --- in this SNR range when number of errors is small and code rate is rather high, pure AWGN channel (even without  quantization) is close to BSC. It is known that BCH codes are good and specially designed for pure BSC channel. And also these codes probably the best short codes with very high code rate. A bit worse results give DE Polar and DE +  ACE  LDPC --- these two codes were  also specially designed for a fixed high SNR values and thus their constructions are channel dependent. Channel independent PW Polar, RM Polar and  PEG LDPC are also close to each other and have a bit worse performance than one for channel-dependent codes. Only TPC codes have  significantly worse performance and can not been considered as code-candidates for 1-bit ADC AWGN channel. 

\begin{figure}[h!]
\centering
\includegraphics[width=\columnwidth,height=150pt]{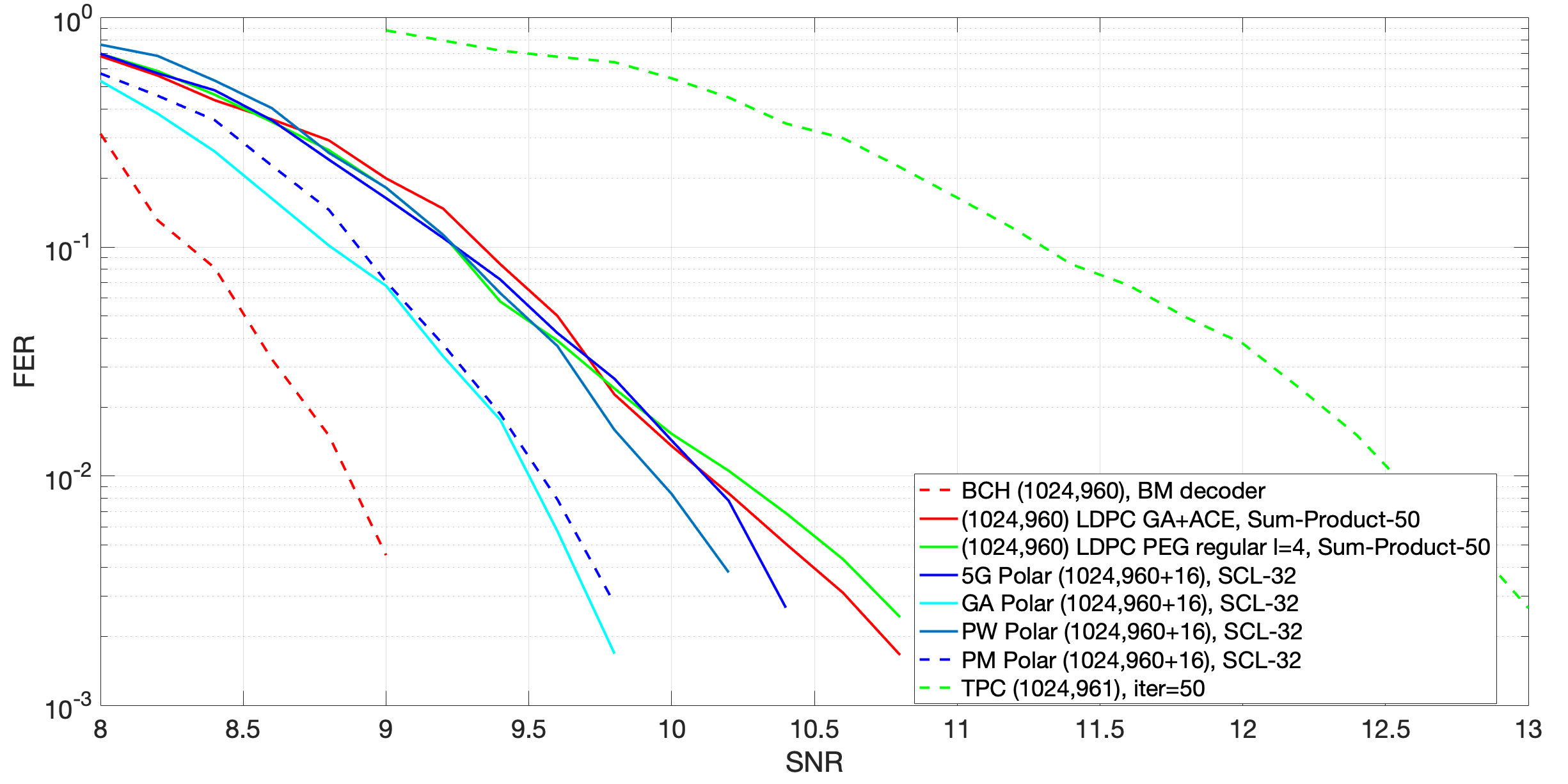}
\caption{FER versus SNR of AWGN+1-bit ADC channel for different code constructions with $N\approx 1024$, $R\approx 0.9375$}
\label{comp09375}
\end{figure}

Simulation results for $N\approx 1024$ and $R\approx 0.9375$ are presented in Fig.~\ref{comp09375}. From this figure we can see that codes can be now divided in four groups: best BCH code, RM polar and DE Polar,  LDPC codes and other polar channel independent codes and the last and worst code is TPC. Instead of the results for $R\in\{0.75,\:0.8125,\:0.875\}$ the difference between all coding schemes become significant. The best performance has BCH code --- in this SNR range when number of errors is small and code rate is rather high, pure AWGN channel can be approximated by BSC with high accuracy. It is known that BCH codes are good and specially designed for pure BSC channel. The next group with a bit worse performance is DE Polar and RM Polar --- DE Polar was specially designed for a fixed high SNR and thus  it  is channel dependent, RM Polar is channel independent but has large minimal distance which is important for these SNR  values. Channel independent PW Polar, 5G Polar,  LDPC  have even worse performance --- all these codes but  DE + ACE LDPC  are  channel independent which is crucial for such high rate. Though DE + ACE LDPC is channel dependent but is well-known that it is rather difficult  to obtain good  LDPC codes for such ultra-high code rate: LDPC  codes have  good performance for medium rates. TPC codes have  significantly worse performance than all other schemes because for this coding rate these codes consist of long single parity check codes (SPC) thus has minimal distance only $d=4$ and can correct only $t=1$ error.

\subsection{Summary}

As a summary observed by simulation results we obtained above, we can make the following conclusions:
\begin{itemize}
    \item LDPC codes and Polar codes have comparable performance at least for short code lengths ($N\approx 1024$).
    \item The higher SNR the better performance of BCH codes. These codes should be used for ultra-high rate transmission scenario ($R\geq 0.8125$).
    \item The more unquantized AWGN channel is closer to BSC channel (when SNR tends to high values) the better performance of BCH codes and minimal distance becomes most important metric in terms of performance. 
    \item The higher SNR the more crucial the channel-dependence of the code construction
    \item The best performance can be obtained by using either DE + ACE LDPC codes or GA Polar codes for moderate SNR and by BCH codes for high SNR
\end{itemize}

\section{Conclusion}

In this paper we compare the performance of different coding schemes for AWGN + 1-bit ADC channel for moderate length and different rates of codes. We gave some recommendations to choose the best code constructions for a given transmission scenario.

\section{ACKNOWLEDGMENT}
This work is an output of a research project implemented as part of the Basic Research Program at the National Research University Higher School of Economics (HSE University).


\begin{thebibliography}{100}

\bibitem{Varasteh} M. Varasteh, B. Rassouli, O. Simeone and D. Gündüz, \emph{Zero-Delay Source-Channel Coding With a Low-Resolution ADC Front End,} in IEEE Transactions on Information Theory, vol. 64, no. 2, pp. 1241-1261, Feb. 2018

 \bibitem{Risi} C. Risi, D. Persson and E. G. Larsson, \emph{Channel estimation and performance analysis of one-bit massive MIMO systems,}
 
\bibitem{Choi} J. Choi, J. Mo and R. W. Heath Jr., \emph{Near maximum-likelihood detector
and channel estimator for uplink multiuser massive MIMO systems with
one-bit ADCs,} IEEE Trans. Commun., vol. 64, no. 5, pp. 2005-2018,
May 2016.
\bibitem{Li} Y. Li, C. Tao, G. Seco-Granados, A. Mezghani, A. L. Swindlehurst, and
L. Liu, \emph{Channel estimation and performance analysis of one-bit massive
MIMO systems}
\bibitem{Muta} E. M. Mohamed, O. Muta, and H. Furukawa, \emph{Static and dynamic 
channel estimation techniques for MIMO-constant envelope 
modulation,} in GLOBECOM Workshops (GC Wkshps), 2011 IEEE, 
2011, pp. 549-554

\bibitem{Mohamed} E. M. Mohamed, \emph{A complexity efficient equalization technique for 
MIMO-constant envelope modulation,} in Wireless Technology and 
Applications (ISWTA), 2013 IEEE Symposium on, 2013, pp. 114-
119



\bibitem{Kim} S. Kim, N. Lee and S. Hong, \emph{Uplink Multiuser Massive MIMO Systems with One-Bit ADCs: A Coding-Theoretic Viewpoint} 2017 IEEE Wireless Communications and Networking Conference (WCNC), 2017

\bibitem{Lee} S. Kim, N. Lee and S. Hong, \emph{Uplink Massive MIMO Systems with One-Bit ADCs: A Low-Complexity Weighted Minimum Distance Decoding,} GLOBECOM 2017 - 2017 IEEE Global Communications Conference, 2017

\bibitem{Cho} Y. Cho, S. Kim and S. Hong, \emph{Successive Cancellation Soft Output Detector for Uplink MU-MIMO Systems with One-Bit ADCs,} 2018 IEEE International Conference on Communications (ICC), 2018

\bibitem{Amrallah} A. Amrallah, H. S. Hussein and E. M. Mohamed, \emph{Comparative study of channel coding techniques for MIMO-CEM system with IF sampled 1-bit ADC,} 2017 Advances in Wireless and Optical Communications (RTUWO), 2017

\bibitem{Alencar} R.R.M. de Alencar,L.T.N. Landau, R.C. de Lamare,  \emph{Continuous phase modulation with 1-bit quantization and oversampling using iterative detection and decoding.} J Wireless Com Network 2020

\bibitem{Balevi} E. Balevi and J. G. Andrews, \emph{High Rate Communication over One-Bit Quantized Channels via Deep Learning and LDPC Codes,} 2020 IEEE 21st International Workshop on Signal Processing Advances in Wireless Communications (SPAWC), 2020
\bibitem{Andrews}  E. Balevi and J. G. Andrews, \emph{Autoencoder-based error correction coding for one-bit quantization}, IEEE Trans. on Communications, 2020.

\bibitem{Singh}
J.~Singh,O.~Dabeer and U.~Madhow, \emph{On the limits of communication with low-precision analog-to-digital conversion at the receiver}, IEEE Transactions on Communications, 2009, Vol. 57, no. 12, pp. 3629---3639.  

\bibitem{Bioglio}
V.~Bioglio, C.~Condo and I.~Land, \emph{Design of polar codes in 5G new radio}, IEEE Communications Surveys and Tutorials. – 2020.

\bibitem{Huawei}
Huawei and HiSilicon. (2016, Aug) \emph{Polar code design and rate matching.} 3GPP meeting.

\bibitem{Mori}	
R.~Mori and T.~Tanaka, \emph{Performance of polar codes with the construction using density evolution}, IEEE Communications Letters, 2009, Vol. 13, no. 7, pp. 519---521.

\bibitem{Trif}	
P.~Trifonov, \emph{Efficient design and decoding of polar codes}, IEEE Transactions on Communications, 2012, Vol. 60, no. 11, pp. 3221---3227.

\bibitem{Arikan}	
E.~Arikan \emph{Channel polarization: A method for constructing capacity-achieving codes for symmetric binary-input memoryless channels}, IEEE Transactions on information Theory, 2009, Vol. 55, no. 7, pp. 3051---3073.

\bibitem{Tal}
I.~Tal and A.~Vardy \emph{List decoding of polar codes}, IEEE Transactions on Information Theory, 2015. Vol. 61, no. 5, pp. 2213---2226.

\bibitem{Gallager}
R.~Gallager, \emph{Low-density parity-check codes}, IRE Transactions on information theory, 1962, Vol. 8, no. 1, pp. 21---28.

\bibitem{Chilappagari}
S. K.~Chilappagari, S.~Sankaranarayanan and B.~Vasic, \emph{Error floors of LDPC codes on the binary symmetric channel}, 2006 IEEE International Conference on Communications, IEEE, 2006, Vol. 3, pp. 1089---1094.

\bibitem{Tanner81}
R.~Tanner, \emph{A recursive approach to low complexity codes}, IEEE Transactions on information theory, 1981, Vol. 27, no. 5, pp. 533---547.

\bibitem{Richardson}
T. J.~Richardson and  R. L.~Urbanke, \emph{The capacity of low-density parity-check codes under message-passing decoding}, IEEE Transactions on information theory, 2001,  Vol. 47, no. 2, pp. 599---618.

\bibitem{ACE}
T.~Tian  et al. \emph{Selective avoidance of cycles in irregular LDPC code construction}, IEEE Transactions on Communications, 2004,  Vol. 52, no. 8, pp. 1242---1247.

\bibitem{Massey} J. ~Massey, \emph{Shift-register synthesis and BCH decoding,} in IEEE Transactions on Information Theory, vol. 15, no. 1, pp. 122-127, January 1960

\bibitem{Elias}
P.~Elias, \emph{Error-free coding}, 1954.

\bibitem{Condo}
V.~Bioglio, C.~Condo and I.~Land, \emph{Construction and decoding of product codes with non-systematic polar codes}, 2019 IEEE Wireless Communications and Networking Conference (WCNC), IEEE, 2019, pp. 1---6.


\end{thebibliography}
\end{document}